\documentclass[traditabstract,rnote]{aa}
\usepackage{natbib} 
\usepackage{graphicx}
\usepackage{txfonts}

\begin{document}

\title{More evidence for extinction of activity in galaxies}

\author{A. Marecki \and B. Swoboda}
\offprints{Andrzej Marecki \email{amr@astro.uni.torun.pl}}

\institute{Toru\'n Centre for Astronomy, N. Copernicus University,
           87-100 Toru\'n, Poland}

\date{Received 4 March 2011 / Accepted 24 March 2011}

\abstract {This {\em Research Note} amends an article in which we showed that 
radio-loud quasars can become radio-quiet. Exploring the analogy between 
galactic nuclei and X-ray binaries (XRB), we pointed out there that this 
transition in quasars could be identified with a switch from low/hard to 
high/soft state in microquasars. Here, we present the evidence that traces 
of past occurrences of this kind of phenomena can be found in normal but 
once active galaxies. Based on the properties of a few such ``post-active'' 
galaxies that are representative for a much wider group, it has been argued
that they have reached the evolutionary stages when their nuclei, which were 
radio-loud in the past, now, mimicking the behaviour of XRBs, remain in the 
intermediate state on their way towards quiescence or even have already 
entered the quiescent state. It follows that the {\em full} evolutionary 
track of XRBs can be mapped onto the evolution of galaxies. The above 
findings are in line with those reported recently for IC\,2497, a galaxy 
that 70,000 years ago or less hosted a quasar but now appears as a normal 
one. This scenario stems from the presence of Hanny's Voorwerp, a nebulous 
object in its vicinity excited by that QSO in the epoch when IC\,2497 was 
active. The post-active galaxies we deal with here are accompanied by extremely
weak and diffuse relic radio lobes that were inflated during their former active
period. These relics can be regarded as radio analogues of Hanny's Voorwerp.}


\maketitle


\section{Introduction}

The amount of evidence that physical properties of all systems containing a 
black hole (BH) are qualitatively analogous and quantitatively scalable with 
the BH mass is substantial -- see \citet{ms11}, hereafter Paper\,I, and 
references therein. Observations of X-ray binaries (XRB), where one of the 
components is a stellar-mass BH, make it clear that as these objects evolve, 
they move on the hardness--intensity diagram (HID) along a characteristic 
path -- see Fig.\,7 in \citet{Fender2004}. However, various changes of the XRBs'
properties expressed in the HID, which can be directly observed, cannot be 
followed for galaxies, whose central BHs are supermassive, because the pace 
of the evolution scales with the BH mass \citep{McHardy2006}. Nevertheless, 
for a large sample of quasars, an analogue of HID called a 
disc-fraction/luminosity diagram (DFLD) was constructed by 
\citet{Koerding2006}. Moreover, the authors transformed the HID of a simulated 
sample of XRBs into a DFLD and it turned out that these two DFLDs were 
similar. This way, an important argument was given in favour of the unification
and scalability of different classes of object centred on a BH.

We showed in Paper\,I that galactic nuclei bearing signatures of a 
transition between fundamentally different areas on DFLD do exist. The three 
QSOs presented there recently left the radio-loud (RL) very high (VH) state 
and moved to the radio-quiet (RQ) high/soft (HS) state\footnote{The
correspondence between radio and X-ray states was comprehensively discussed
in Paper\,I (Sect. 4.2) and references therein.}. It must be noted, 
though, that the closed turtle-head-like loops microquasars make in their 
HIDs are not fully reproduced in DFLDs of quasars. While it is conspicuous 
that microquasars traverse the HIDs from the HS state to the quiescent state 
through the intermediate state \citep{Fender2004}, the respective transition 
for galactic nuclei has not been identified so far. This can be easily 
explained as a selection effect. \citet{Koerding2006} could not show these 
evolutionary stages in galactic DFLDs for an elementary reason: their study was 
limited to quasars. Therefore, if we postulate that the analogy between 
galactic nuclei and XRBs is valid for any evolutionary stage, a gradual
cessation of the {\em whole} activity is a transition that also has to be
allowed for. It follows that a reconstruction of the lower part of the ``turtle 
head'' that represents the intermediate and quiescent states in HID could be 
possible for a DFLD of galaxies if both active and ``post-active'' ones were 
taken into account.

The question that arises at this point is whether any signatures of a putative 
post-active state in galaxies do exist. What we claim is that in some 
circumstances radio lobes that straddle galaxies can be used as such a 
signature. Our approach is based on the fact that the lobes of an RL active 
galaxy are huge reservoirs of energy, so even if relativistic particles from 
the central engine are no longer supplied to the hotspots, a ``coasting'' 
radio source is observable for a substantial amount of time, according to 
\citet{KG1994} -- up to $10^8$\,yr. When the energy supply from the central 
engine to the lobes is cut off eventually, whatever the reason, they become 
weak and diffuse because of lack of fuelling and have steep spectra caused by 
radiation and expansion losses. However, hotspots fade out much sooner -- 
their lifetimes are roughly $7\times 10^4$\,yr \citep{Kaiser2000}, hence 
they are normally not seen in coasting lobes, unless shortly after the 
switch-off of the ``central engine''.

We explored the latter possibility in Paper\,I where we dealt with 
asymmetric radio sources. What we meant by ``asymmetry'' was that one lobe 
was diffuse and devoid of a hotspot, whereas the other resembled a 
Fanaroff-Riley \citep{FR74} type-II (FR\,II) lobe. Assuming that the double 
did not lie in the sky plane, we attributed the asymmetry to the time lag 
between their images: the lobe perceived as a relic was nearer than the lobe 
with a hotspot and consequently it was observed in a later stage of the decay.
We concluded that in asymmetric radio sources the transition from RL/VH state 
to RQ/HS state must have been relatively recent. Here, we focus on 
double-lobed radio sources where both lobes are already diffuse and 
deprived of hotspots, and we treat them as a record of their host galaxies' 
former -- albeit not recent -- activity in the radio domain.


\begin{table}
\caption{Post-active galaxies with extremely diffuse radio lobes}
\label{gal}
\centering
\begin{tabular}{l l c c}
\hline
\hline
\multicolumn{1}{c}{RA} & \multicolumn{1}{c}{Dec} & SDDS objID & $z$ \\
\hline
02 18 12.6   & $-$06 43 36    & --- & --- \\
06 57 51.0   & +63 07 22    & --- & --- \\
07 06 05.1   & +41 08 34    & --- & --- \\
07 32 35.742 & +29 15 18.22 & 587728906634068534 & --- \\
07 34 11.7   & +56 11 20    & --- & --- \\
08 04 27.181 & +16 33 29.28 & 587739376155033971 & 0.1821 \\
08 26 00.035 & +52 27 20.84 & 587725552272540161 & --- \\
08 38 56.240 & +53 51 24.67 & 587725980689957085 & --- \\
08 41 57.628 & +08 29 08.79 & 587735348014809268 & --- \\
08 48 45.721 & +52 46 01.50 & 588007005238198462 & --- \\
09 08 05.811 & +06 06 15.05 & 587734690350825753 & 0.1562 \\
09 13 54.2   & $-$03 32 11    & --- & --- \\
09 26 31.503 & +41 28 59.75 & 587732153641533669 & --- \\
09 37 23.015 & +24 23 39.41 & 587741533848732133 & --- \\
09 46 25.891 & +07 26 53.60 & 587732771039936709 & 0.2162 \\
10 07 24.923 & +28 15 57.66 & 587741391575777399 & --- \\
10 09 09.383 & +04 00 33.35 & 588010359074128174 & --- \\
10 12 36.563 & +40 31 07.16 & 588017626950336695 & 0.1527 \\
10 18 31.487 & +46 09 16.58 & 587732483814523034 & 0.1655 \\
10 22 08.142 & +28 53 41.67 & 587739647277400476 & --- \\
10 24 40.227 & +17 54 05.22 & 587742568396030656 & --- \\
10 29 38.730 & +35 59 30.19 & 587738946674163916 & 0.3591 \\
10 41 19.975 & +52 00 47.26 & 587733080269717630 & --- \\
10 52 00.5   & $-$05 06 40    & --- & --- \\
11 05 48.671 & +23 16 12.91 & 587741828583981602 & --- \\
11 24 35.858 & +49 03 25.91 & 587732483282632923 & --- \\
11 30 09.511 & +09 25 36.01 & 587732771051208963 & --- \\
11 37 49.165 & $-$04 01 44.32 & 587745406825136371 & --- \\
11 44 52.749 & +19 51 31.24 & 587742572687917459 & --- \\
12 01 59.391 & +61 37 00.44 & 588009370690322598 & --- \\
12 11 29.075 & +09 58 56.33 & 587734892754370676 & --- \\
12 14 31.828 & +31 40 59.65 & 587739608090083479 & 0.2025 \\
12 28 19.599 & +44 59 08.44 & 588017626692911383 & --- \\
12 49 39.545 & +02 34 37.34 & 587726032790880371 & --- \\
12 58 44.546 & $-$02 26 27.94 & 587725040635216083 & 0.3692 \\
13 05 26.776 & +36 55 21.04 & 587739097530368104 & 0.1414 \\
13 25 19.251 & +05 21 23.24 & 588010879842123967 & 0.1749 \\
13 30 57.337 & +35 16 50.30 & 587739130804043919 & 0.3158 \\
14 07 08.478 & +49 36 47.79 & 588018056185708804 & 0.3703 \\
14 09 32.223 & +07 39 43.64 & 587730023862174005 & --- \\
14 14 40.4   & +51 17 43    & --- & --- \\
14 15 42.9   & +15 53 59    & --- & --- \\
14 16 46.229 & +11 04 11.68 & 587736478126571640 & --- \\
14 40 15.4   & $-$06 53 38    & --- & --- \\
14 45 32.180 & +35 02 53.56 & 588017979979530362 & 0.1774 \\
14 54 46.049 & +10 15 12.10 & 587736478667506843 & --- \\
15 08 49.952 & +06 09 34.49 & 587730023331791162 & 0.3169 \\
15 26 30.287 & +27 24 53.23 & 587736975809380662 & --- \\
15 27 19.440 & +10 50 15.44 & 587736813669712247 & 0.2268 \\
15 39 58.409 & +48 35 37.77 & 587733604798955838 & --- \\
15 47 46.701 & +14 27 10.12 & 587742061631570383 & --- \\
15 57 04.679 & +34 03 09.47 & 587736781995376949 & 0.3771 \\
16 07 51.319 & +12 40 37.90 & 587742645163917343 & 0.2746 \\
16 15 38.863 & +26 35 38.26 & 587736919434068304 & --- \\
16 16 22.713 & +48 23 31.34 & 588007005265985838 & --- \\
16 25 22.460 & +55 56 35.37 & 587739862026289647 & --- \\
16 31 43.736 & +42 18 09.15 & 587729653958377716 & --- \\
16 36 35.872 & +22 04 08.48 & 587736619865866624 & --- \\
16 47 22.191 & +12 34 07.11 & 587739707425882649 & --- \\
16 53 00.860 & +45 40 23.75 & 587736980642791824 & --- \\
17 05 05.0   & +27 09 58    & --- & --- \\
17 08 18.248 & +34 14 35.18 & 587729782273475192 & --- \\
17 25 17.2   & +39 25 03    & --- & --- \\
17 37 46.865 & +62 39 20.49 & 588011501529072058 & --- \\
21 21 05.297 & +01 08 12.22 & 587731174919504271 & --- \\
21 42 56.497 & +00 27 59.74 & 587730847966692469 & --- \\
21 54 23.503 & +00 05 03.71 & 587734304875741352 & 0.1480 \\
\hline
\end{tabular}
\small{Coordinates are from SDSS if available, otherwise from NVSS.}
\end{table}

\begin{table*}
\caption{Quasar-like objects for which both radio lobes are relics.}
\label{qso}
\label{quasars}
\centering
\begin{tabular}{c c c c c}
\hline
\hline
Object & $z$ & $\log(M_{BH}/M_{\sun})$ & $\log(L_{bol})$ & $L_{bol}/L_{Edd}$\\
\hline
SDSS\,J082905.01+571541.6 & 0.3505 & 9.341 & 45.212 & 0.0059 \\
SDSS\,J123915.40+531414.6 & 0.2013 & 9.051 & 45.121 & 0.0095 \\
\hline
\end{tabular}
\end{table*}


\section{The method and the results}

The method used in this work was the same as that described in Section 2 of 
Paper\,I. It relies upon the fundamental property of the images restored 
from interferometric data: the poorer the coverage of the $u-v$ plane with 
short baselines, the worse the rendering of extended and diffuse areas of the 
image lacking compact features. Specifically, the Very Large Array (VLA) in 
B-conf. used when carrying out the {\it Faint Images of the Radio Sky at 
Twenty-Centimeters} (FIRST) survey \citep{White97} fails to show lobes well 
if they are diffuse, although their presence can be firmly established via 
{\it NRAO/VLA Sky Survey} (NVSS) \citep{Condon1998}. Quantitatively, there 
is a large discrepancy between the peak and the integrated flux of a 
lobe if it is of a relic nature. If there is extreme diffuseness, the relic 
lobes remain undetected in FIRST, while they are evident in NVSS.

For the purpose of the investigation reported in Paper\,I, we devised an 
algorithm calculating the ratio of peak-to-integrated flux densities for 
FIRST catalogue items pertinent to each of the two NVSS lobes. When that 
ratio was very low for all the FIRST components within the given NVSS lobe, 
which meant that the lobe was featureless and therefore likely to be diffuse,
it was regarded as a potential relic. In a number of cases, the NVSS lobe had 
no FIRST counterpart and, obviously, we treated these as true positives. 
Application of the above recipe by means of an automated procedure, later 
refined by visual inspection of the FIRST maps, led to a selection of 182 
targets with at least one lobe being clearly diffuse and deprived of a 
hotspot.

Whilst Paper\,I was focused on the objects with one relic and one ``active'' 
lobe, having radio cores, and identified with QSOs, we now looked for 
targets without radio cores\footnote{The criterion was that there should be 
no item in the FIRST catalogue in the optical position of the galactic 
nucleus.} and with both lobes decaying. Sixty-nine objects of this kind were 
found in the parent sample consisting of 182. Only two of them -- 
\object{SDSS\,J082905.01+571541.6} and \object{SDSS\,J123915.40+531414.6} -- 
were classified as QSOs and as such listed in the quasar catalogue based on 
Sloan Digital Sky Survey (SDSS) Data Release\,7 \citep{DR7QSO}. Some of 
their physical data taken from \citet{Shen2010} are quoted in 
Table\,\ref{qso}. Their NVSS and FIRST radio maps along with SDSS spectra 
are shown in Figs.\,\ref{082904+571541} and\,\ref{123915+531416}, 
respectively. Although broad and narrow emission lines typical for QSOs are 
present there, the continua, unlike in QSOs, do not rise towards the blue end 
of the spectra. This makes labelling these two objects as quasars 
questionable. Instead, they should be named radio galaxies, but the 
diffuse nature of their radio lobes casts doubt on this classification as well.

The remaining 67 objects are listed in Table\,\ref{gal}. Given that on the 
one hand they are all double lobed in NVSS maps but on the other their lobes 
are extremely diffuse as seen -- or better yet: hardly or even not seen -- 
in FIRST, they could be labelled post-active. Eleven out of those 67 are not 
included in SDSS, therefore their (approximate) coordinates shown in 
Table\,\ref{gal} we extracted from NVSS. The remaining 56 are 
itemised in SDSS but spectra have been provided for only eighteen. Redshifts 
calculated from those spectra are listed in column\,4 of Table\,\ref{gal}. 
To date, no other redshift information pertinent to the galaxies from 
Table\,\ref{gal} is available in the NASA/IPAC Extragalactic Database.


\section{Interpretation of the observational material}

\begin{figure*}
\centering 
\includegraphics[width=7.5cm]{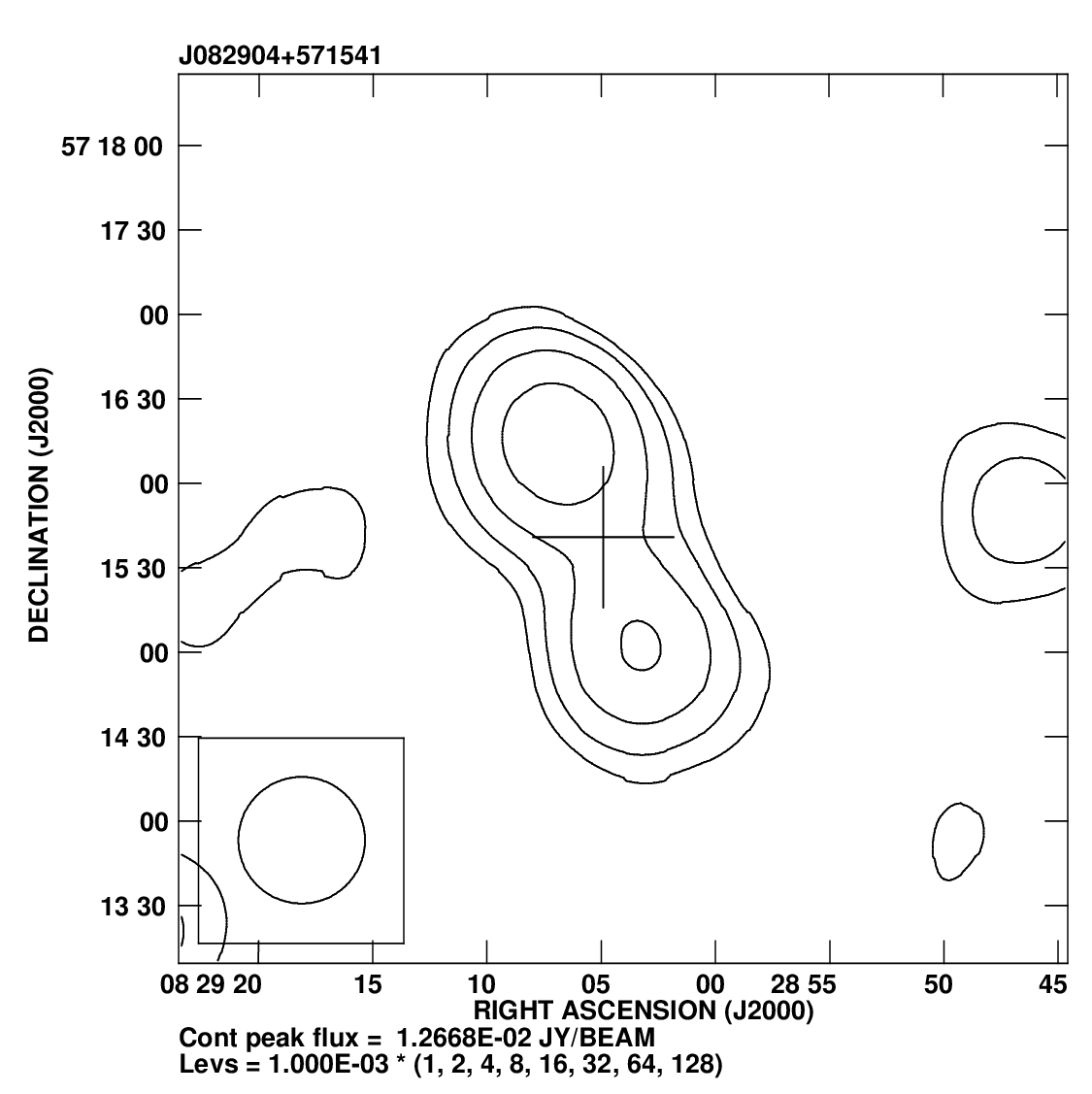} 
\includegraphics[width=7.5cm]{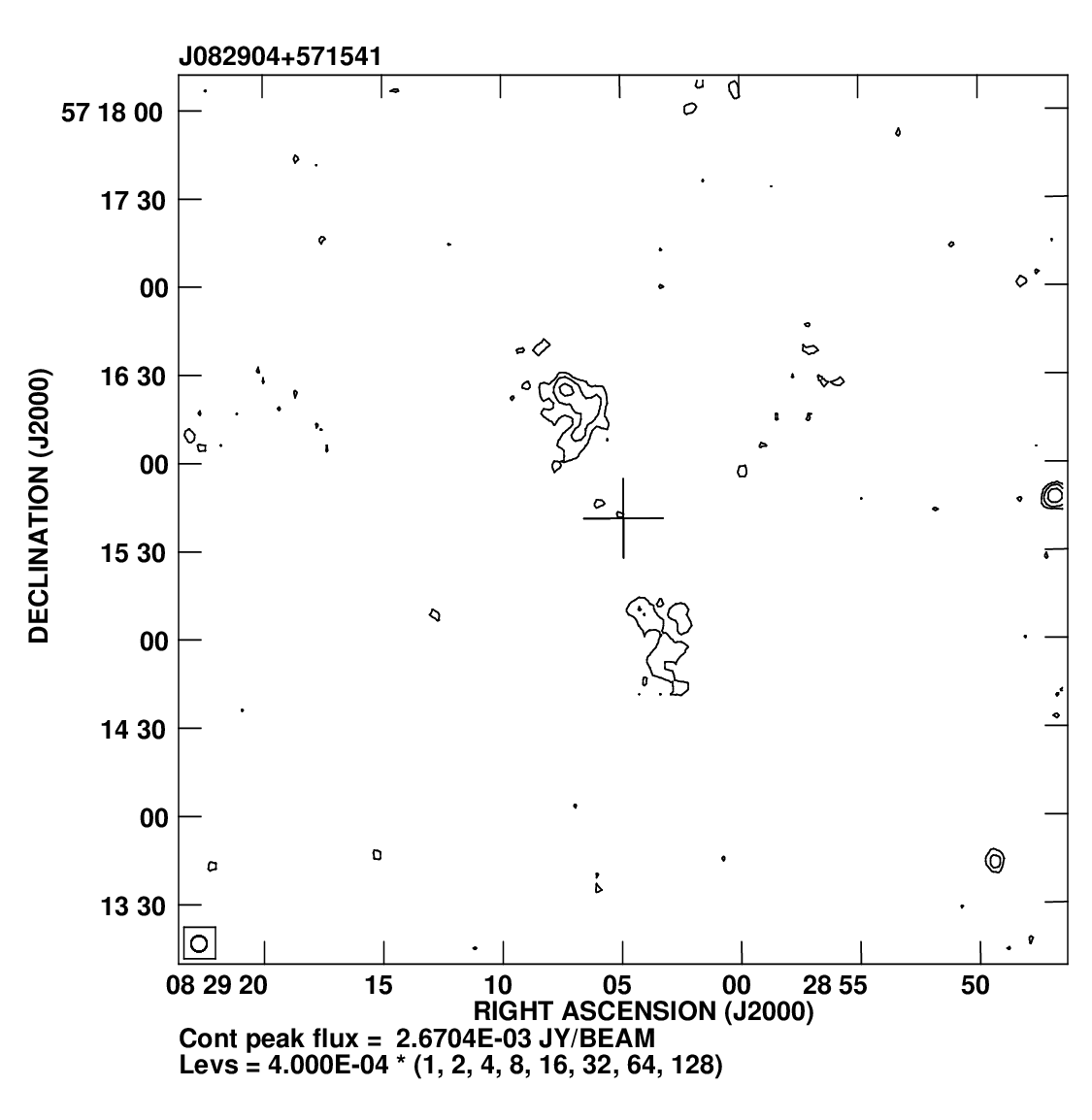}

\vspace{2cm}
\includegraphics[width=12cm,angle=-90]{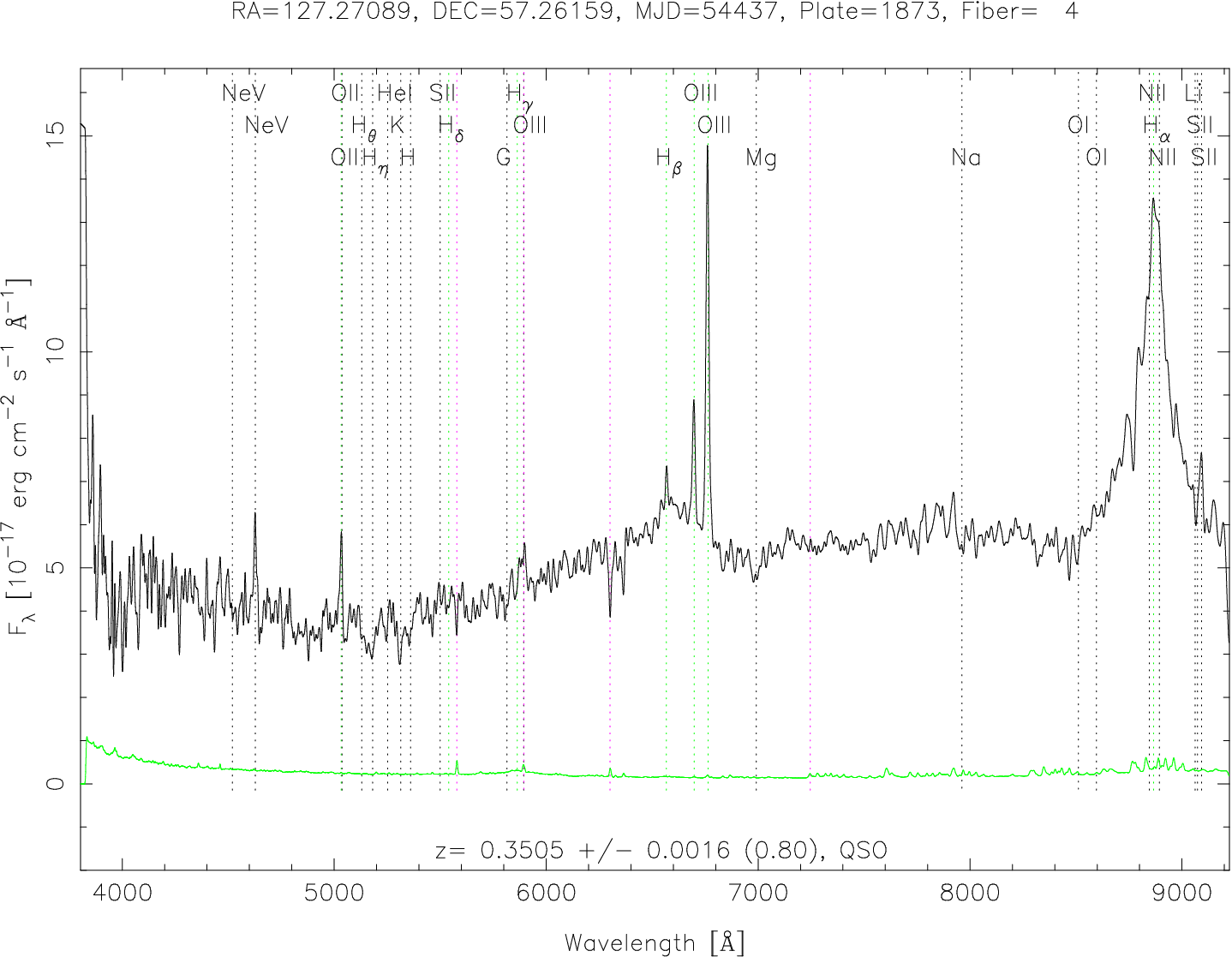}
\caption{NVSS (upper left) and FIRST (upper right) images of
SDSS\,J082905.01+571541.6. Contours are increased by a factor of 2;
the first contour level corresponds to 1\,mJy/beam for NVSS image and 
0.4\,mJy/beam for the FIRST image. Crosses indicate the position of the
galactic nucleus. Lower panel: SDSS spectrum of SDSS\,J082905.01+571541.6.}
\label{082904+571541}
\end{figure*}

\begin{figure*}
\centering 
\includegraphics[width=7.5cm]{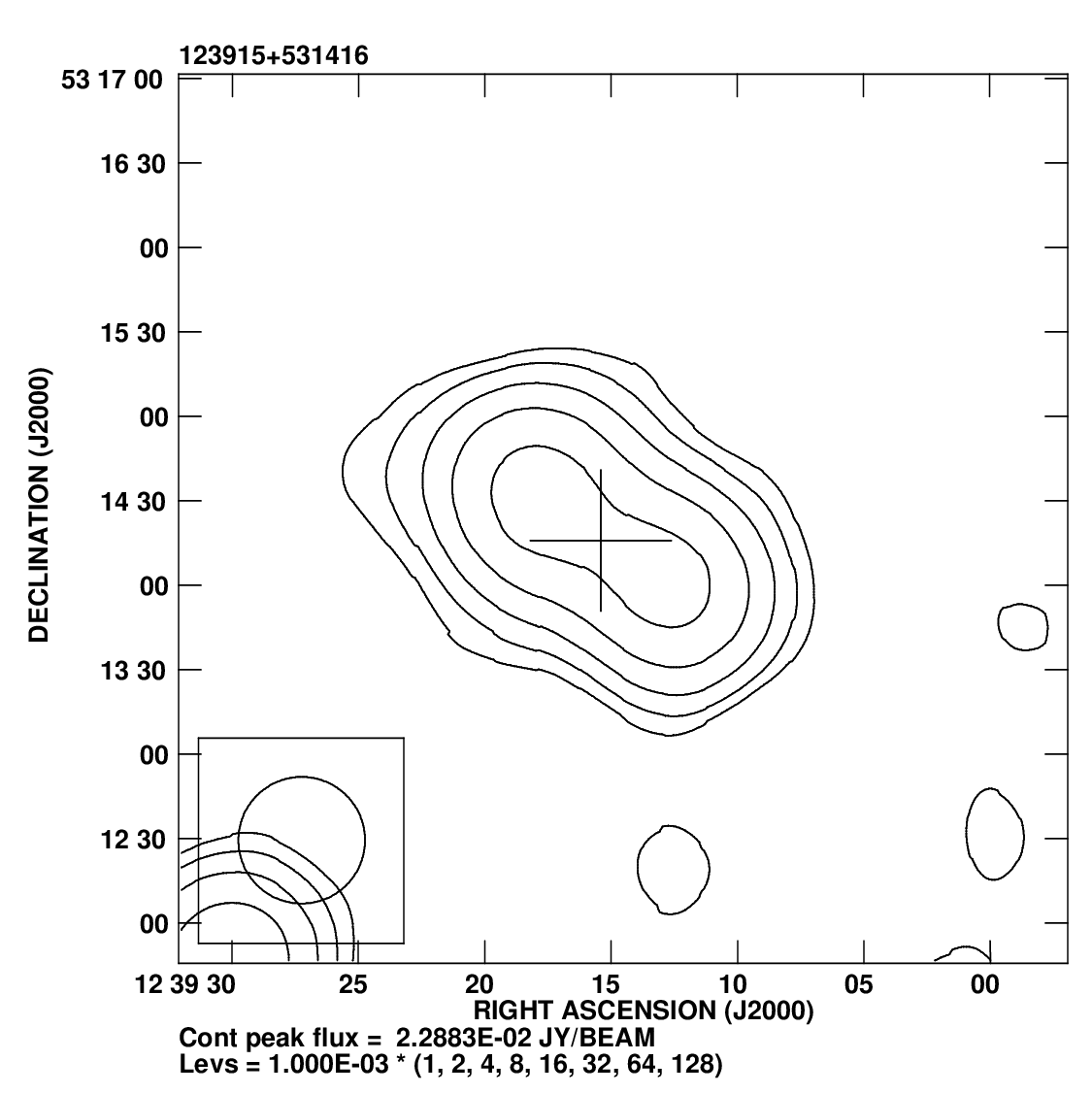} 
\includegraphics[width=7.5cm]{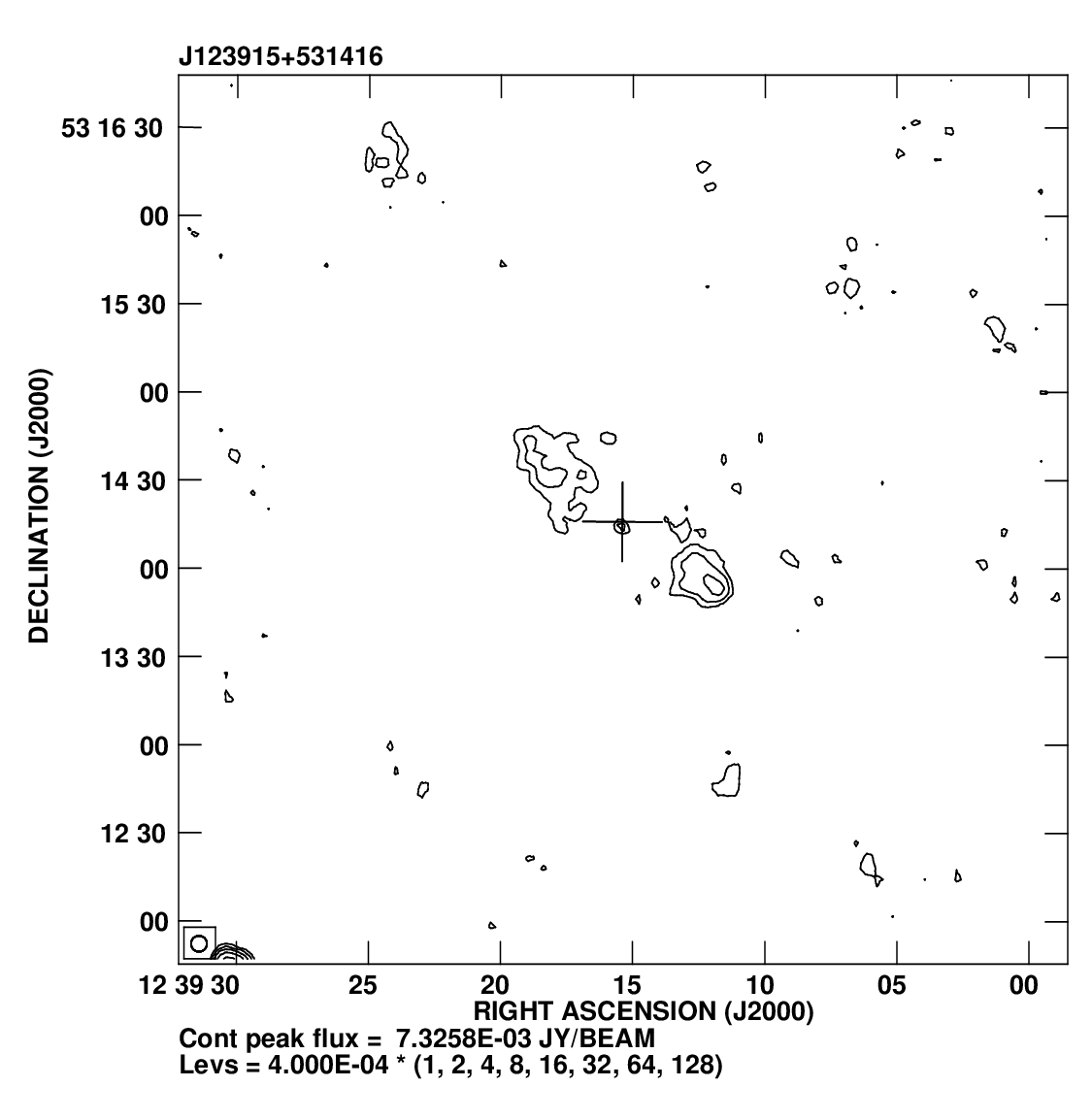}

\vspace{2cm}
\includegraphics[width=12cm,angle=-90]{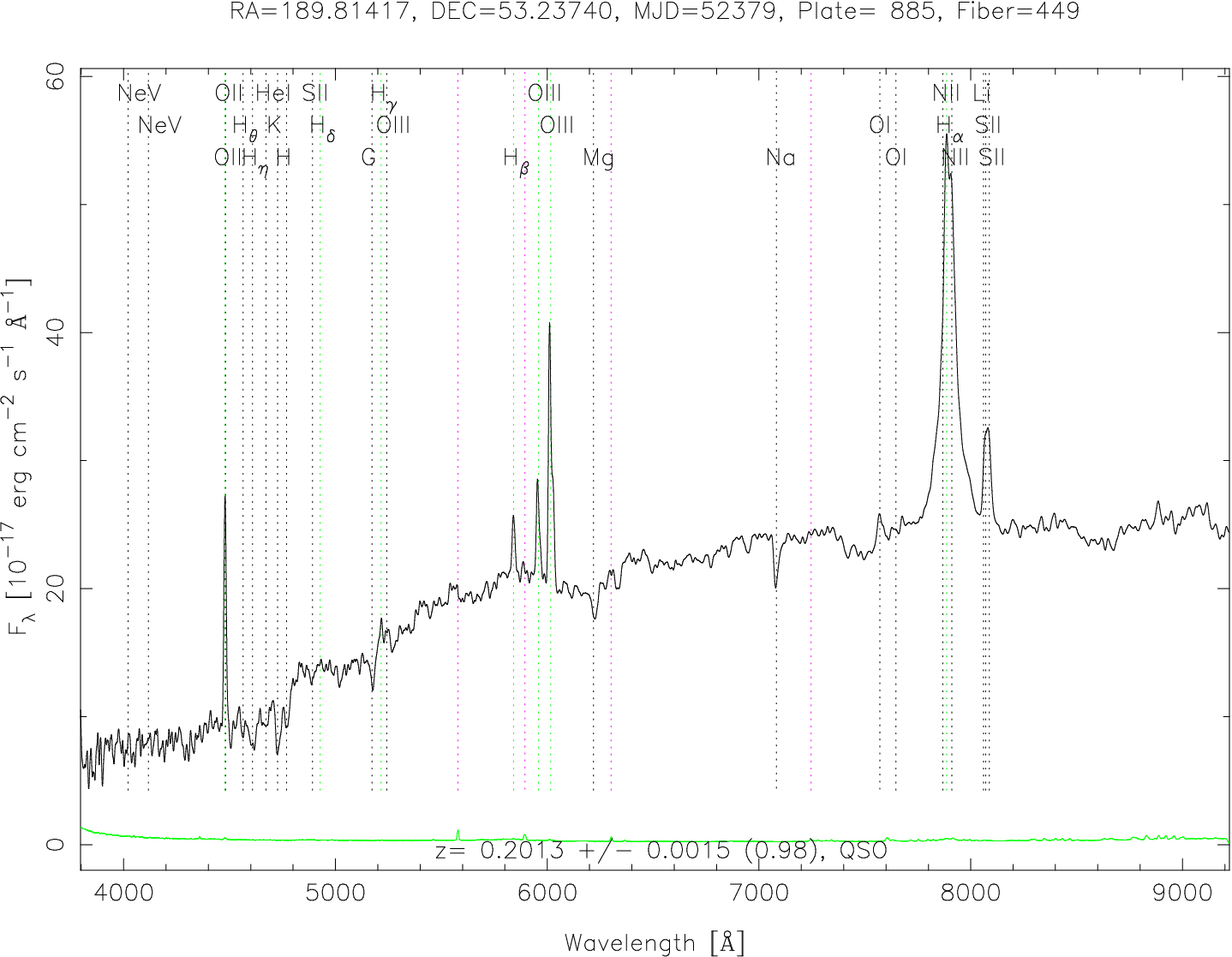}
\caption{NVSS (upper left) and FIRST (upper right) images of
SDSS\,J123915.40+531414.6. Contours are increased by a factor of 2;
the first contour level corresponds to 1\,mJy/beam for NVSS image and 
0.4\,mJy/beam for the FIRST image. Crosses indicate the position of the
galactic nucleus. Lower panel: SDSS spectrum of SDSS\,J123915.40+531414.6.}
\label{123915+531416}
\end{figure*}

Along with the results presented in Paper\,I, this work is an endeavour to 
identify the evolutionary track of galaxies in which the nuclear activity 
switched off. Originally, they were RL, i.e., they remained in a VH state. When 
their activity in the radio domain -- but only in {\em this} domain -- had 
ceased, they moved to the HS state. Three objects of that kind were shown in 
Paper\,I. Its outcome and related considerations in the literature quoted
therein provide a sound basis for a notion of a common evolutionary path
of XRBs and galactic nuclei. It implies that galactic nuclei, once active, can 
leave the HS state and, like XRBs, enter the intermediate state on their way to 
quiescence, which means they become post-active. Objects shown in 
Figs.\,\ref{082904+571541} and\,\ref{123915+531416} are representative for 
the initial stage of such a transition. Although officially labelled QSOs 
\citep{DR7QSO,Shen2010} and even though the emission lines are still present 
in their spectra, neither the continua nor very low Eddington ratios -- see 
the last column of Table\,\ref{qso} -- resemble those of {\em bona fide} 
quasars. Most intriguing here, however, are radio lobes straddling each of 
these two: being diffuse and devoid of hotspots, they are clearly in the 
coasting phase. As mentioned in Sect.\,1, this can be noticeably long -- up to 
$10^8$\,yr \citep{KG1994} -- so assuming that formerly RL galactic nuclei 
could enter, stay, and leave HS state all within that time, a reminiscence 
of the past RL episode in the form of relic radio lobes could indeed be 
observed.

\begin{figure*}
\centering 
\includegraphics[width=7.5cm]{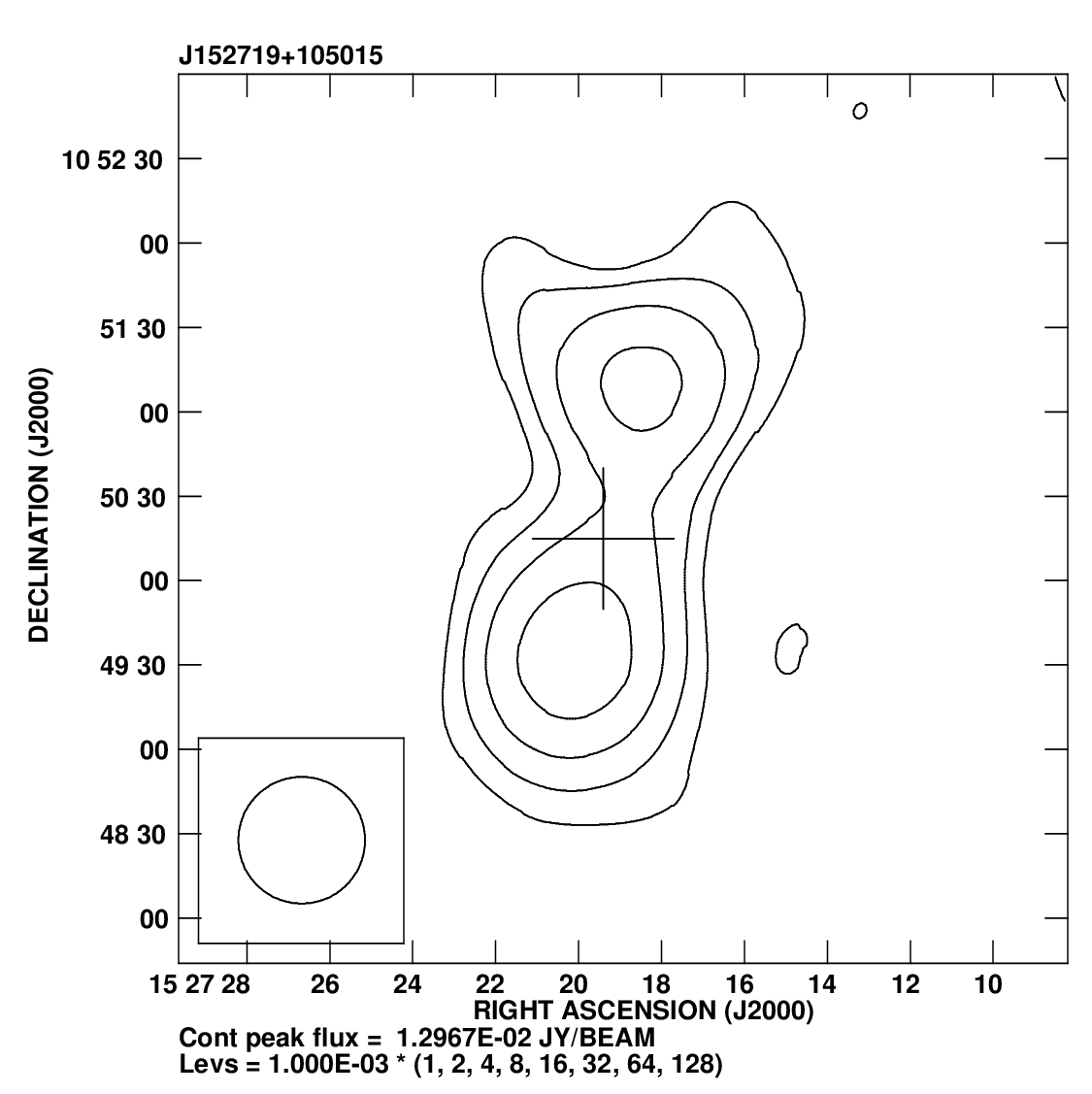} 
\includegraphics[width=7.5cm]{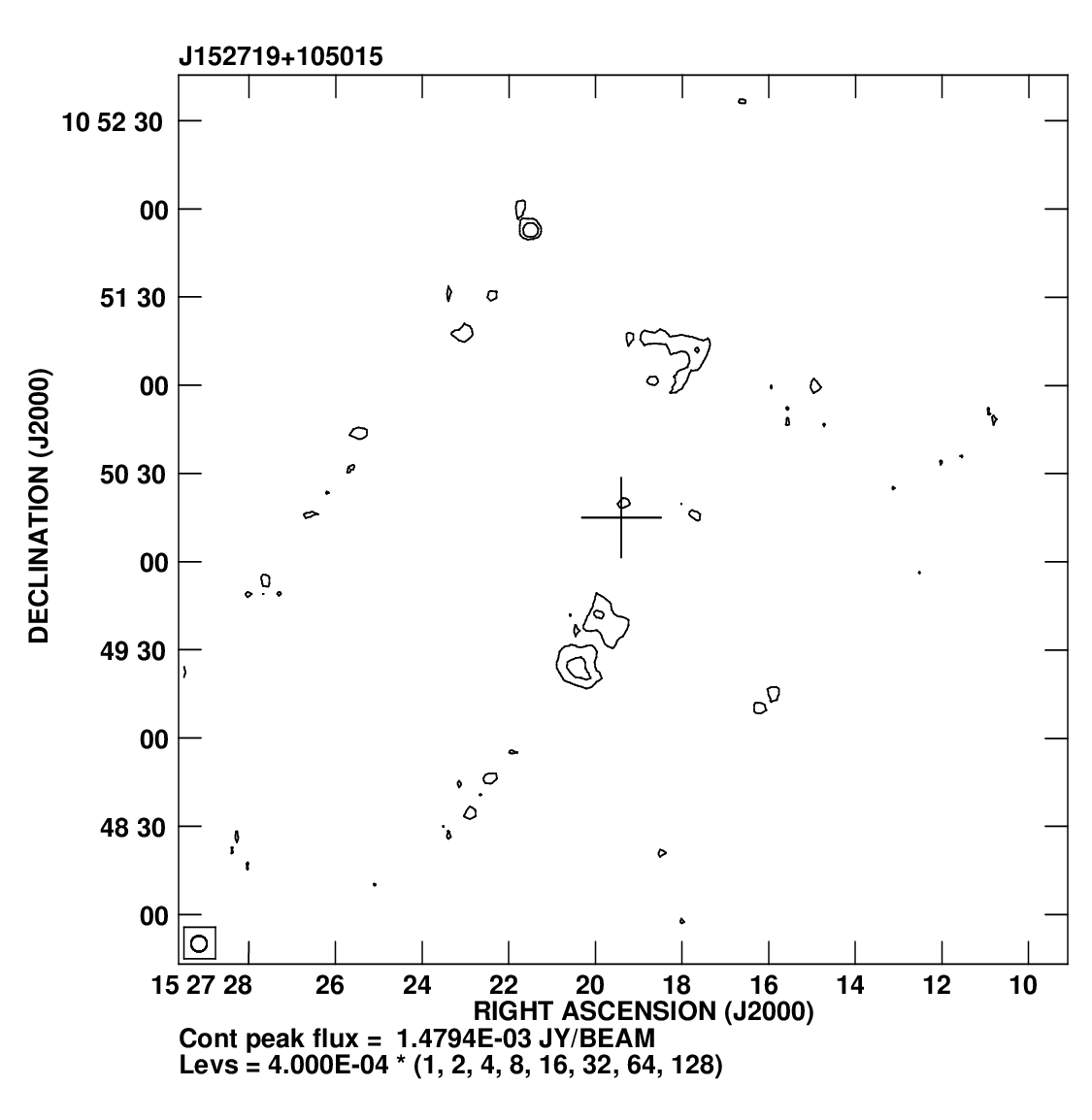} 

\vspace{2cm}
\includegraphics[width=12cm,angle=-90]{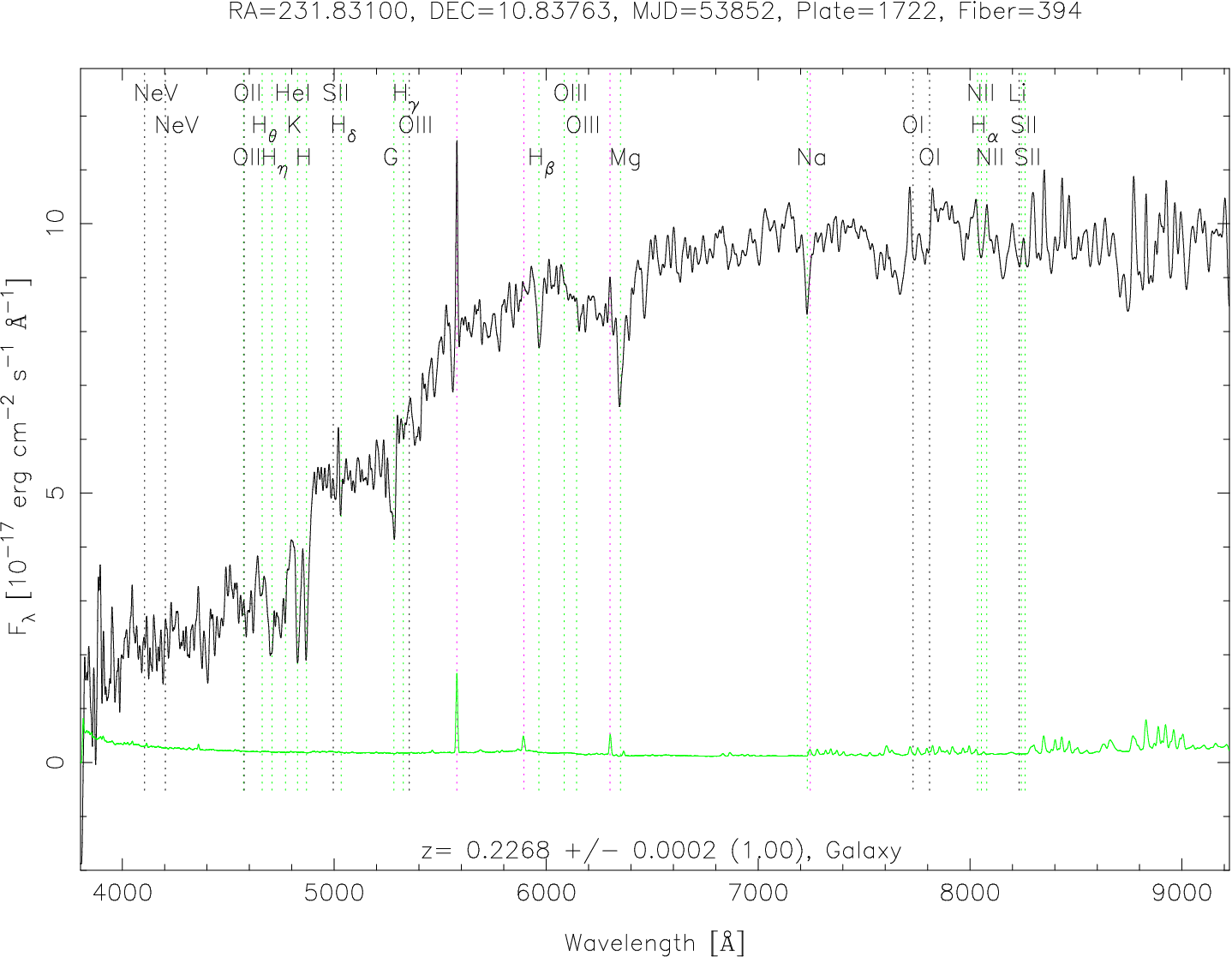}
\caption{NVSS (upper left) and FIRST (upper right) images of 
\object{SDSS\,J152719.44+105015.4}. Contours are increased by a factor of 2; 
the first contour level corresponds to 1\,mJy/beam for NVSS image and 
0.4\,mJy/beam for the FIRST image. Crosses indicate the position of the
galactic nucleus. Lower panel: SDSS spectrum of SDSS\,J152719.44+105015.4.}
\label{152719+105015}
\end{figure*}

\begin{figure*}
\centering 
\includegraphics[width=7.5cm]{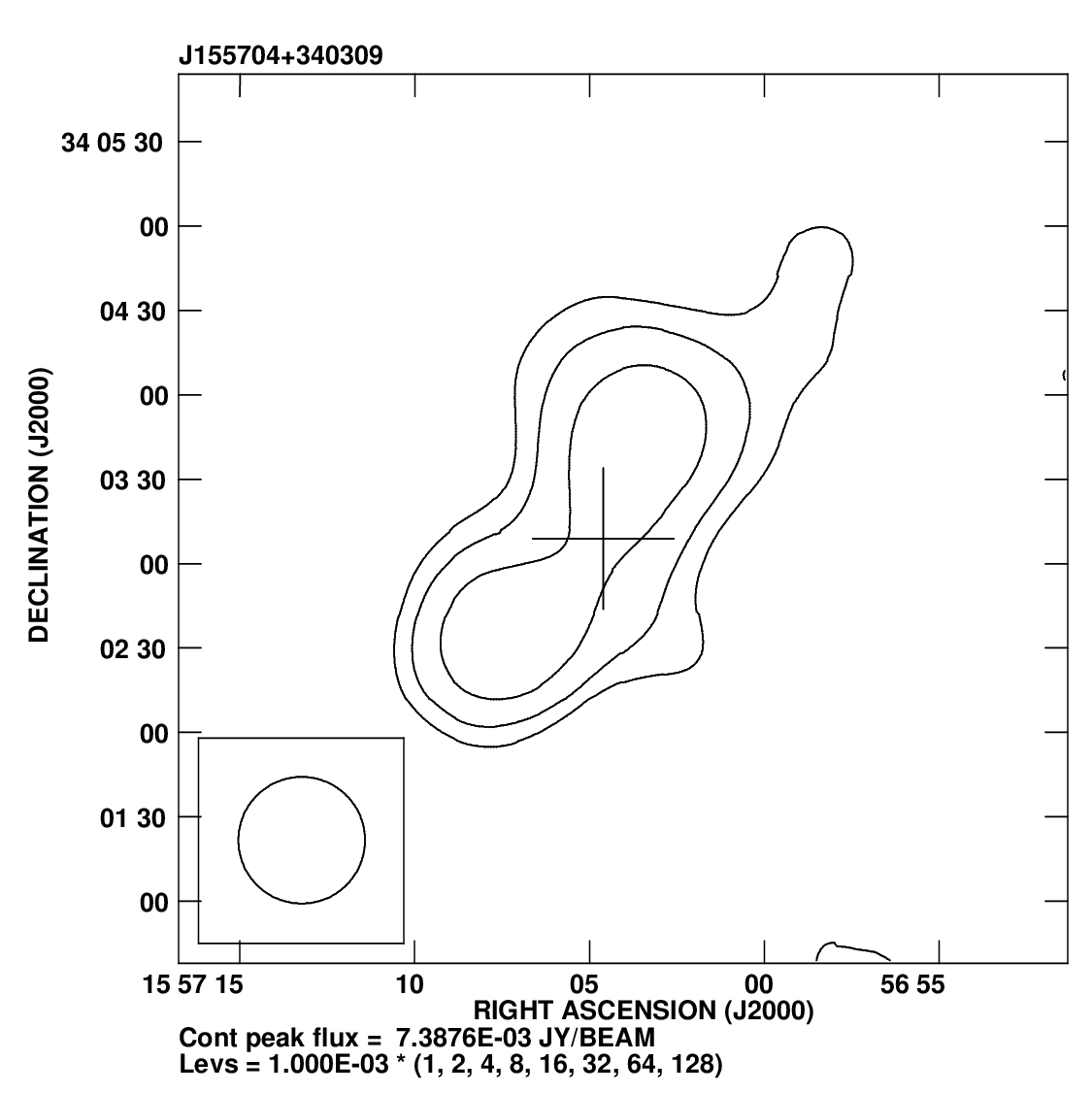} 
\includegraphics[width=7.5cm]{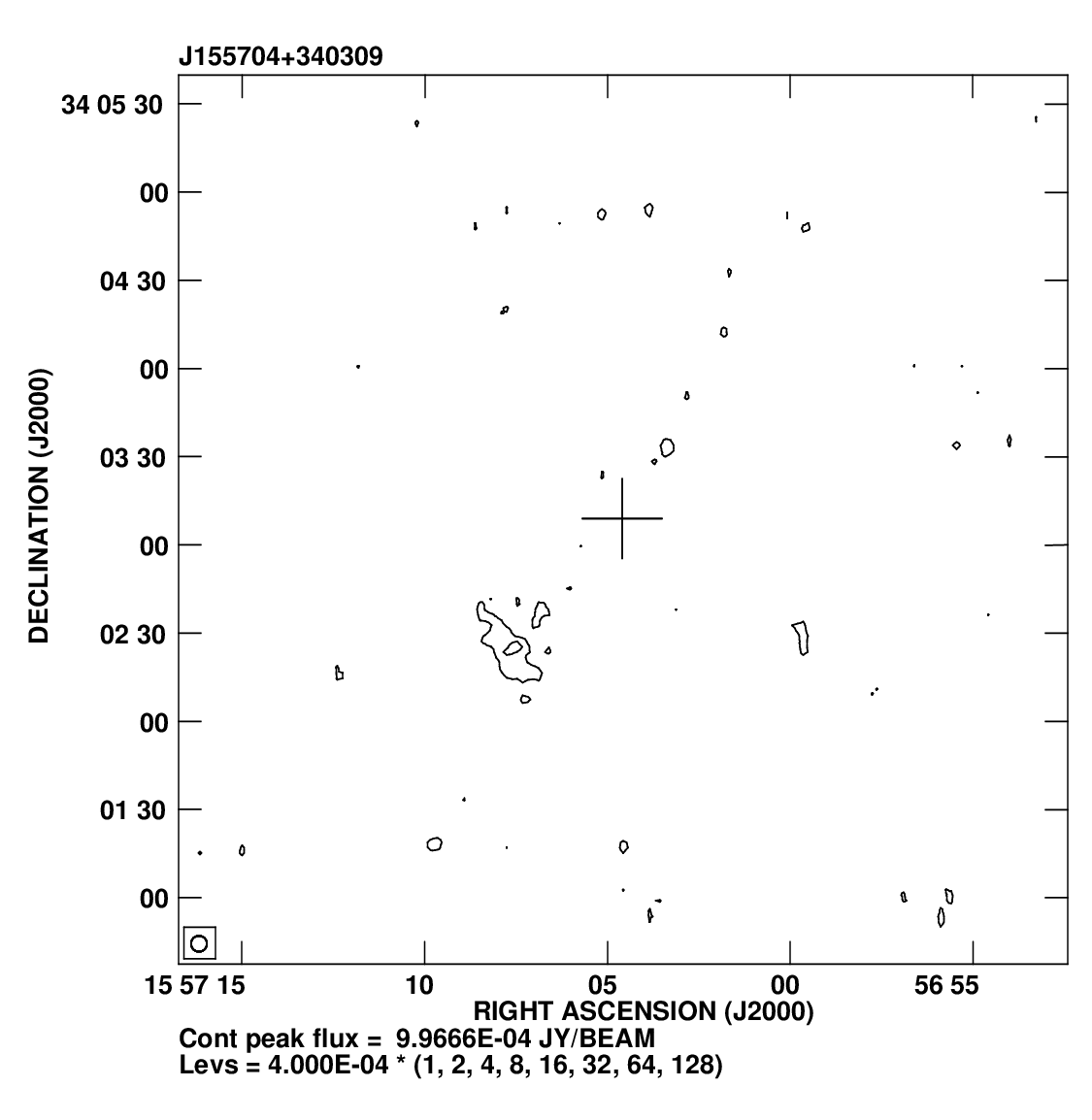}

\vspace{2cm}
\includegraphics[width=12cm,angle=-90]{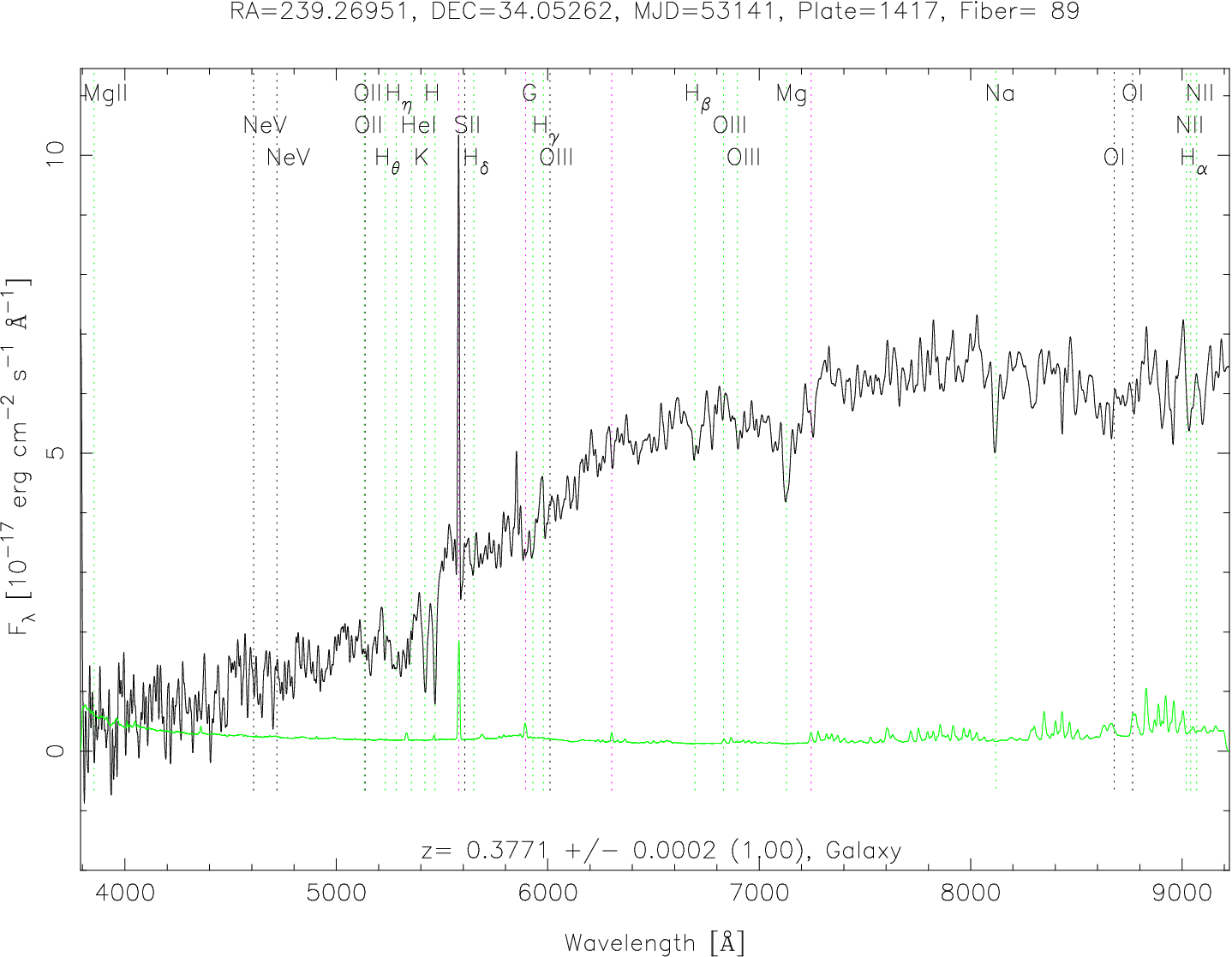}
\caption{NVSS (upper left) and FIRST (upper right) images of
\object{SDSS\,J155704.67+340309.4}. Contours are increased by a factor of 2; 
the first contour level corresponds to 1\,mJy/beam for NVSS image and 
0.4\,mJy/beam for the FIRST image. Crosses indicate the position of the
galactic nucleus. Lower panel: SDSS spectrum of SDSS\,J155704.67+340309.4.}
\label{155704+340309}
\end{figure*}

\begin{figure*}
\centering 
\includegraphics[width=7.5cm]{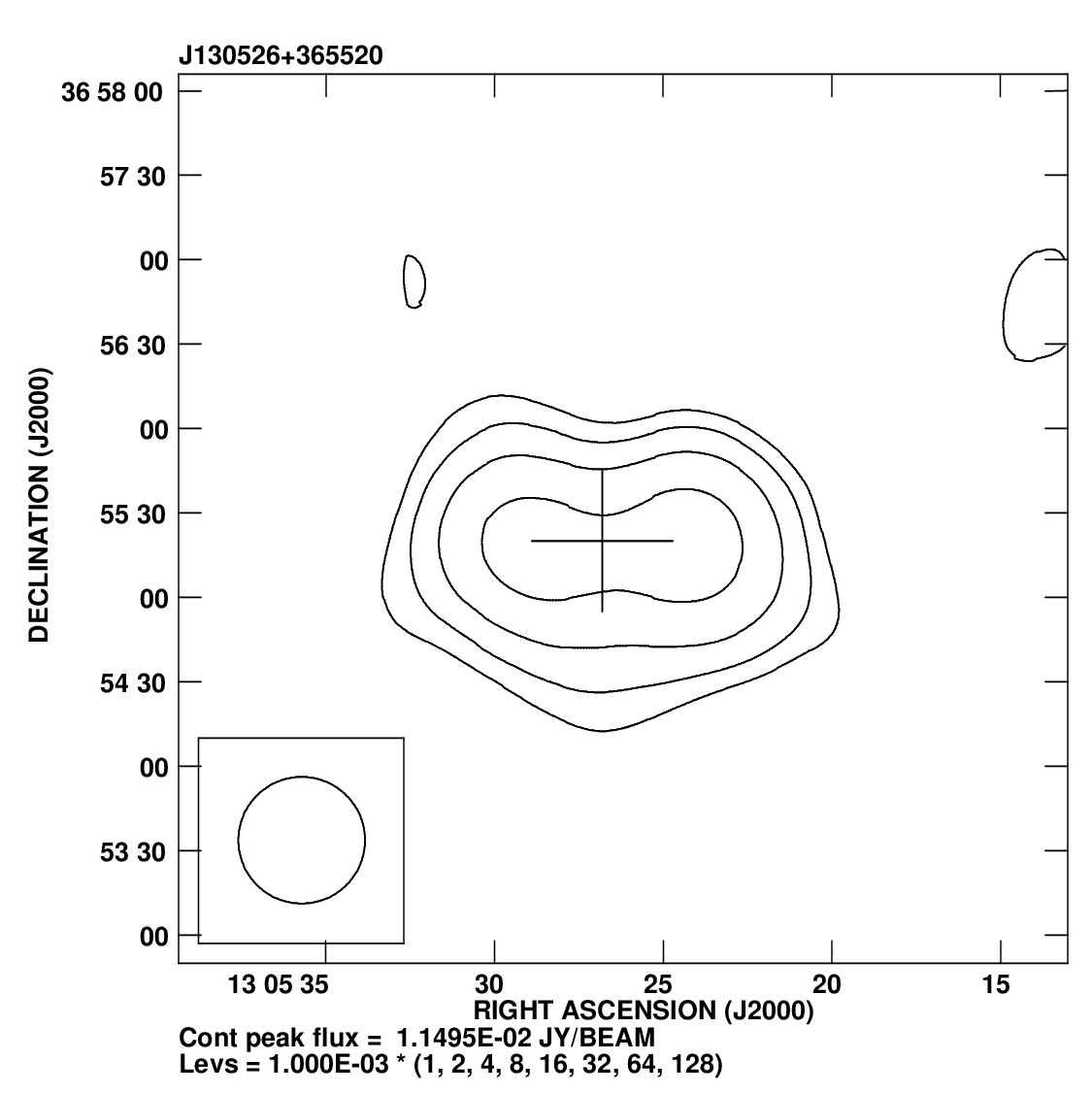} 
\includegraphics[width=7.5cm]{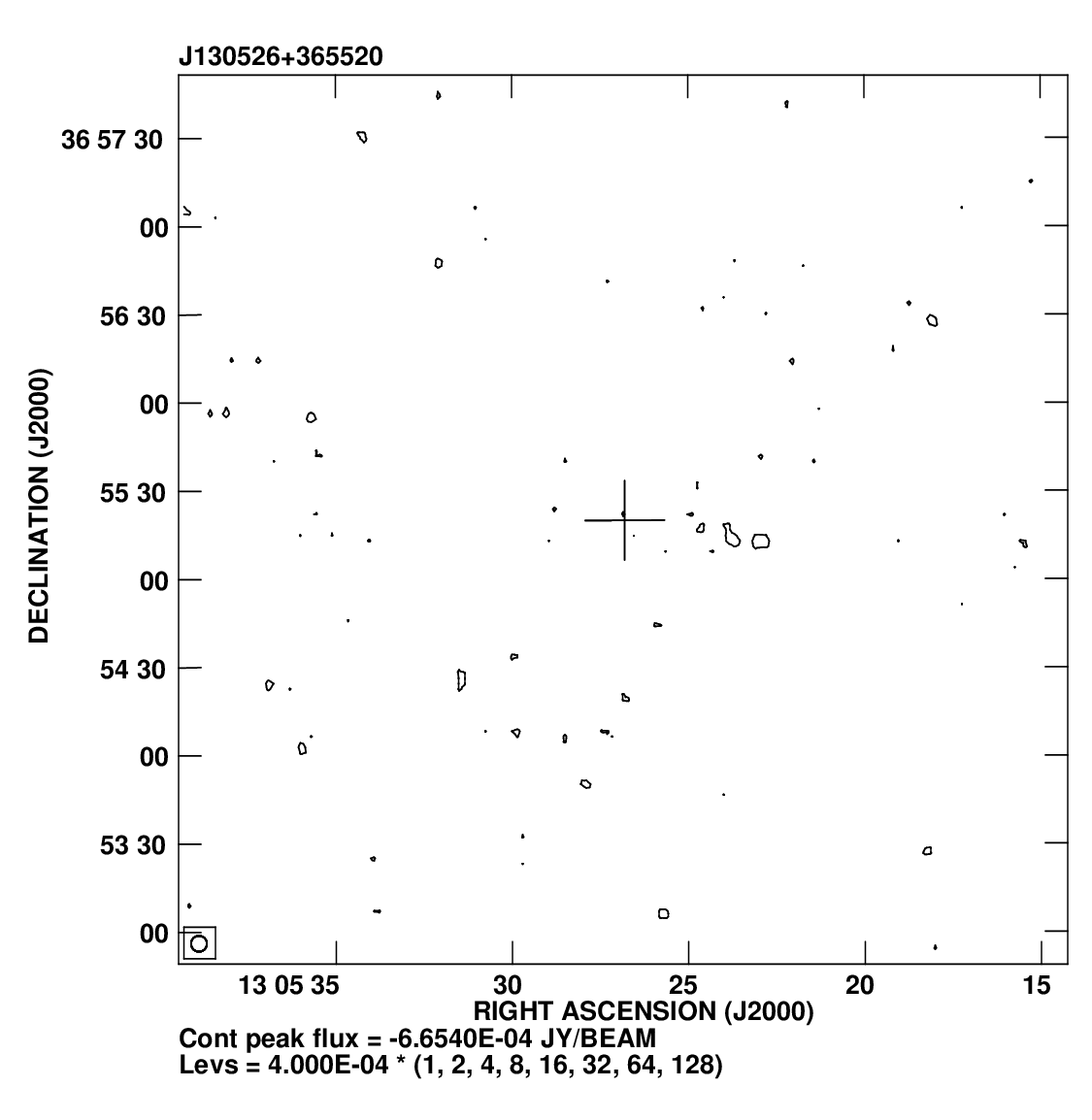}

\vspace{2cm}
\includegraphics[width=12cm,angle=-90]{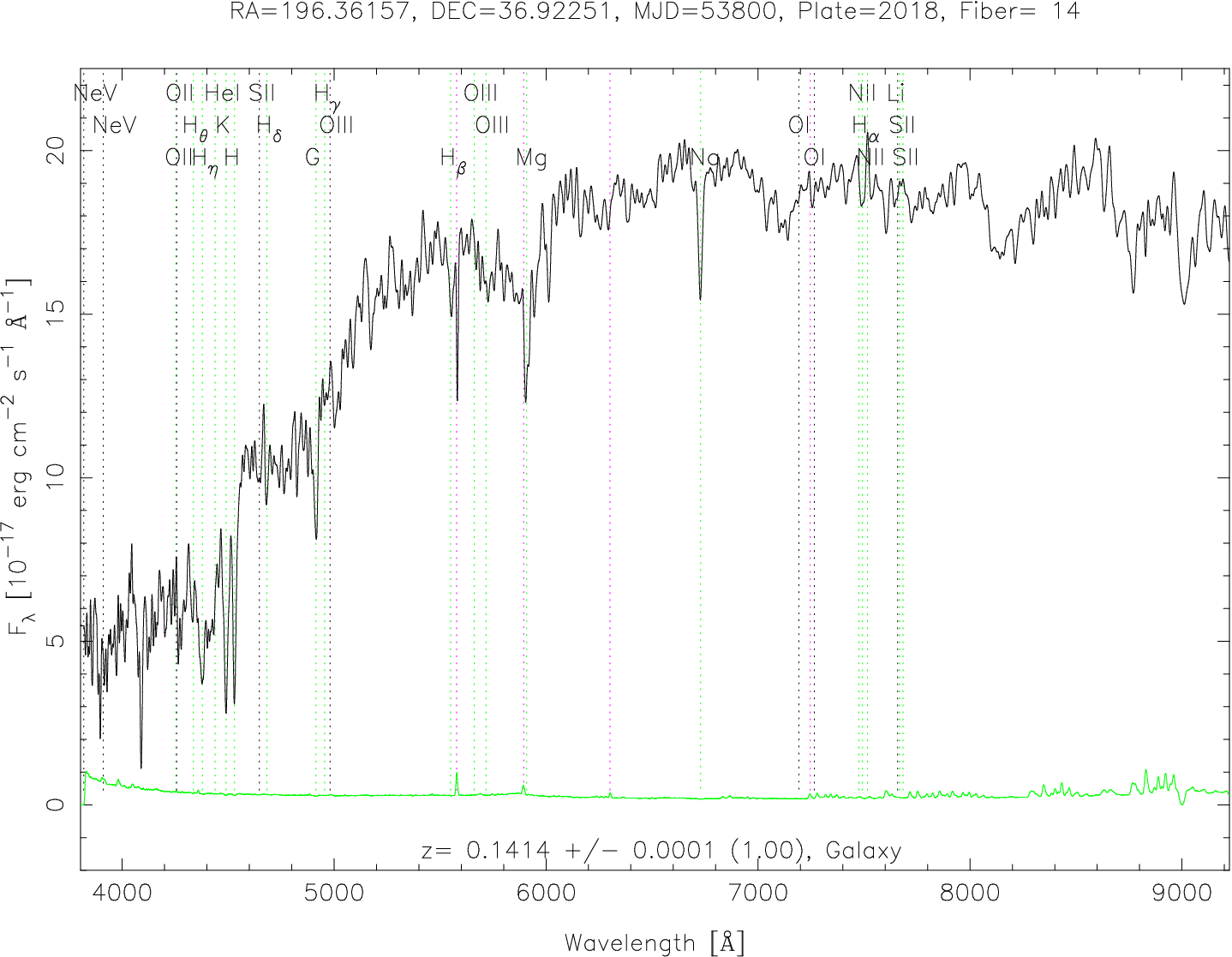}
\caption{NVSS (upper left) and FIRST (upper right) images of
\object{SDSS\,J130526.77+365521.0}. Contours are increased by a factor of 2; 
the first contour level corresponds to 1\,mJy/beam for NVSS image and 
0.4\,mJy/beam for the FIRST image. Crosses indicate the position of the
galactic nucleus. Lower panel: SDSS spectrum of SDSS\,J130526.77+365521.0.
Based on the appearance of this spectrum, SDSS\,J130526.77+365521.0 is a normal 
galaxy, although it must have been active and RL before.}
\label{130526+365520}
\end{figure*}

\begin{figure*}
\centering 
\includegraphics[width=7.5cm]{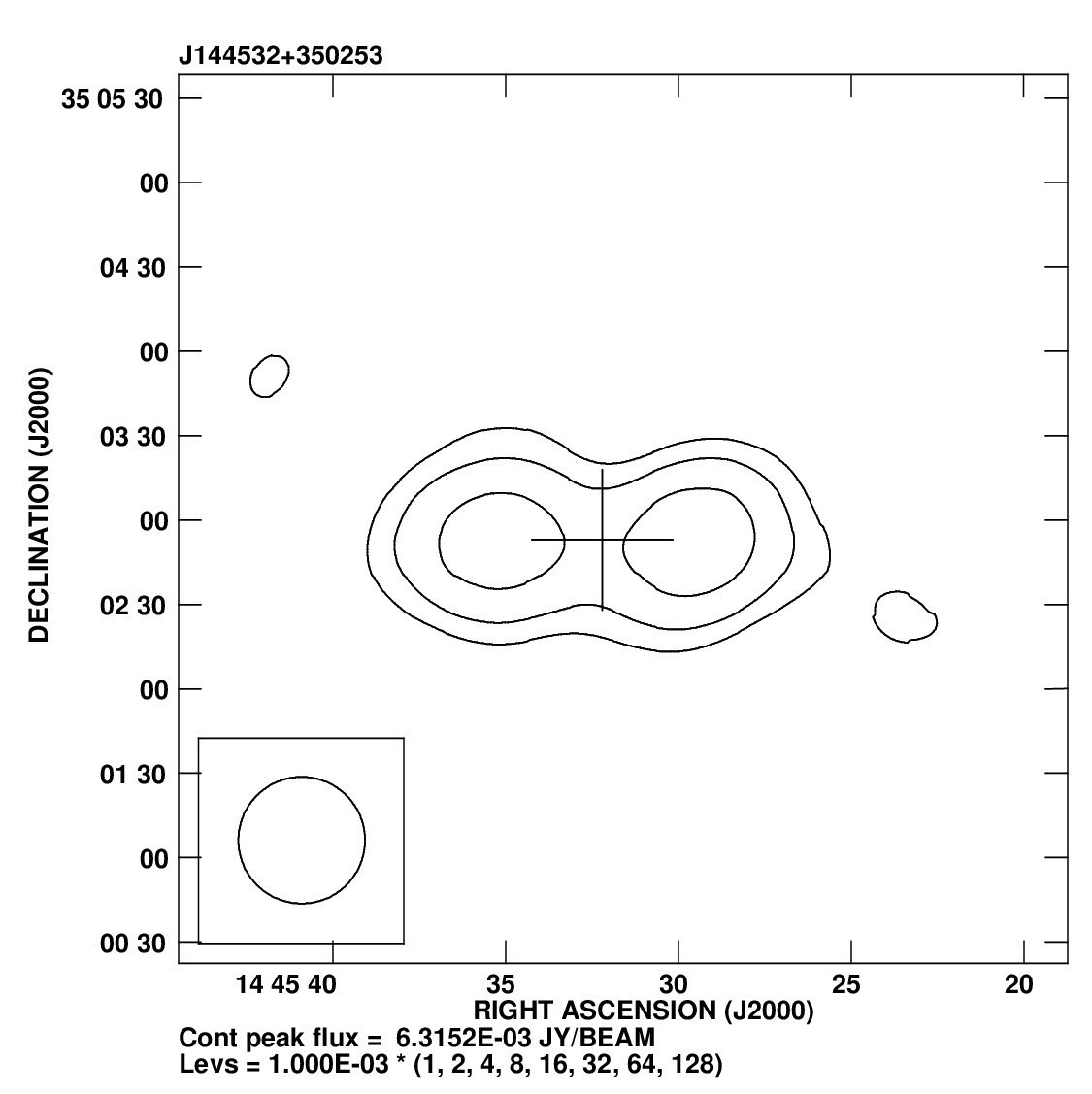} 
\includegraphics[width=7.5cm]{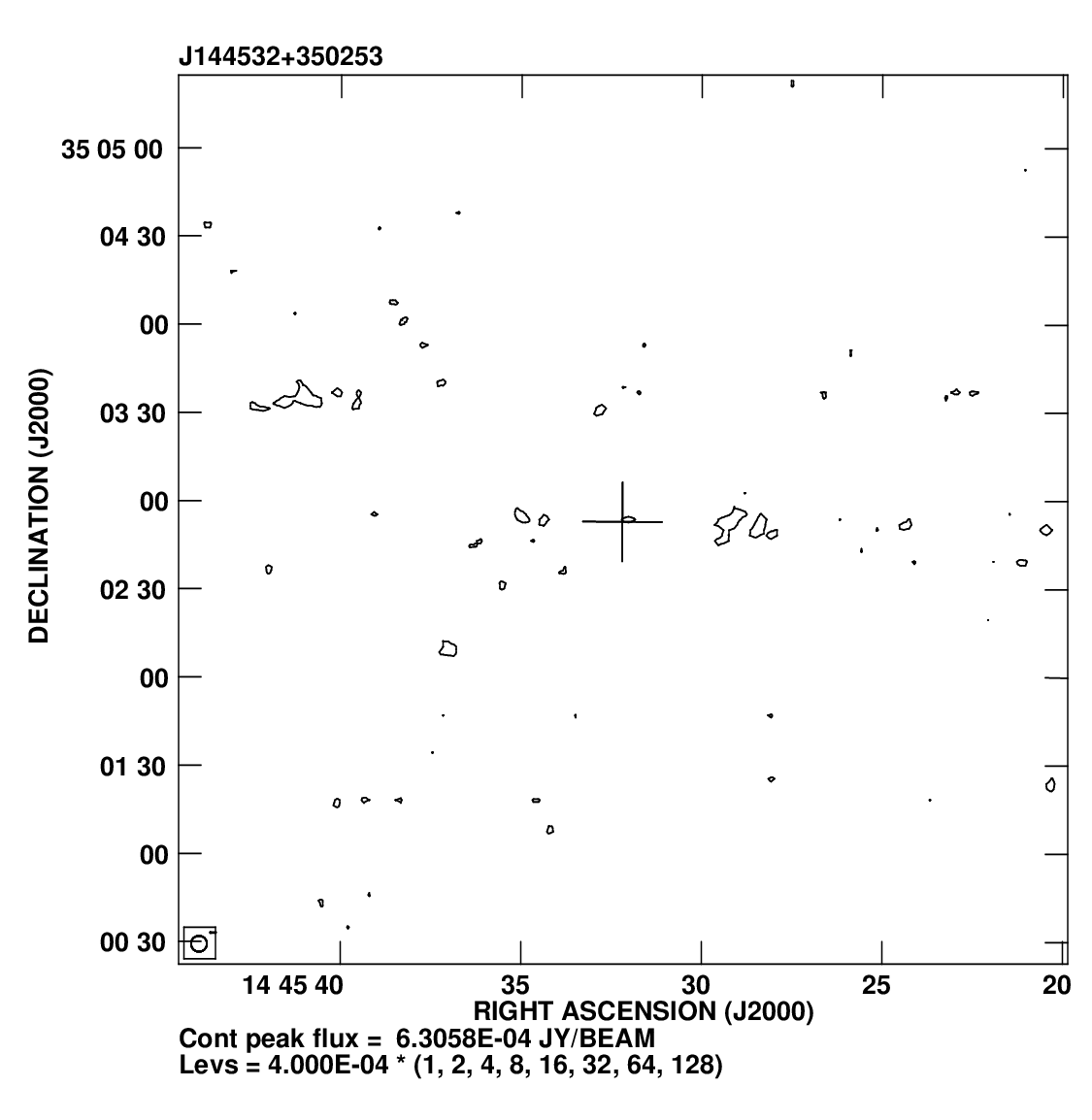}

\vspace{2cm}
\includegraphics[width=12cm,angle=-90]{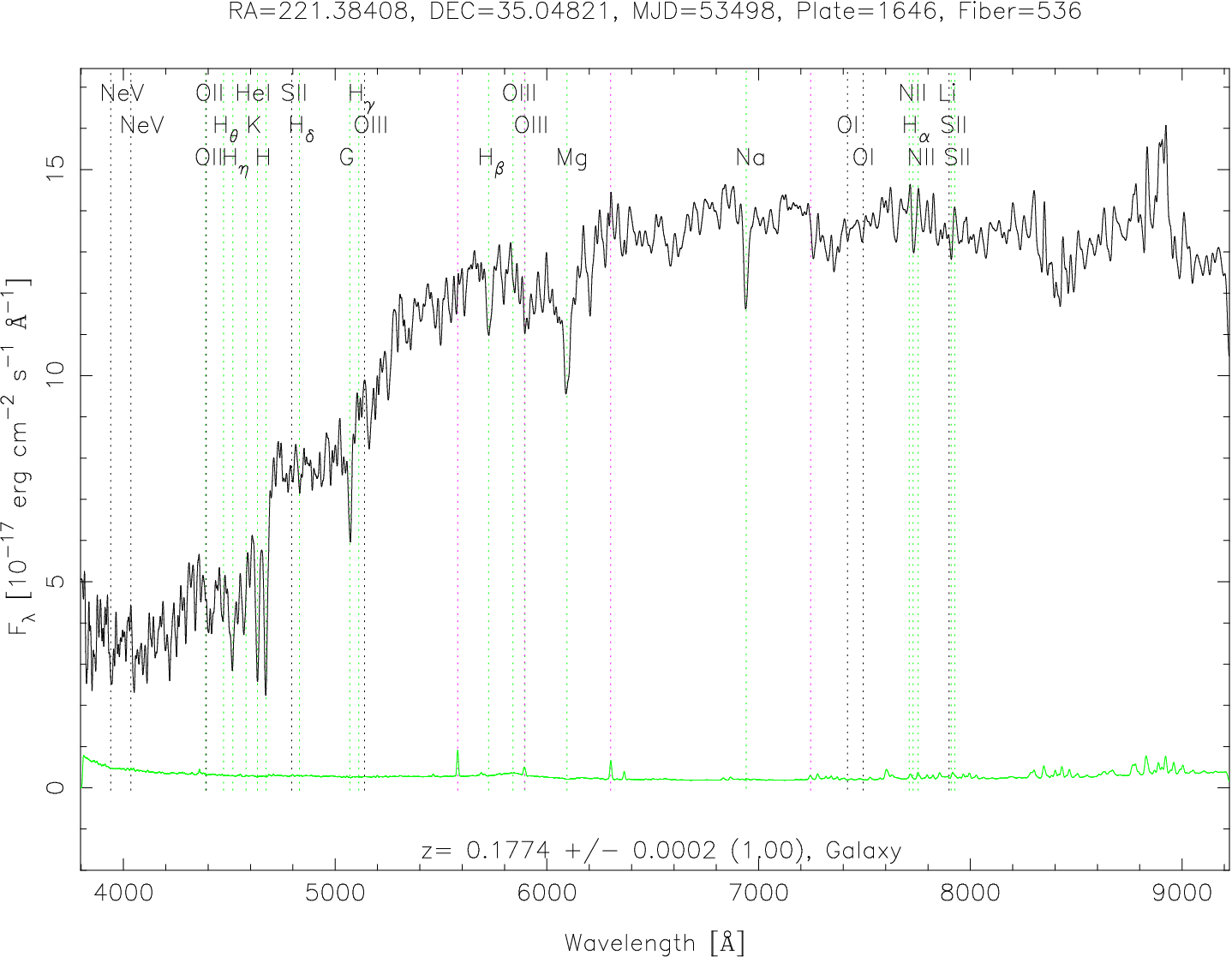}
\caption{NVSS (upper left) and FIRST (upper right) images of
\object{SDSS\,J144532.17+350253.5}. Contours are increased by a factor of 2; 
the first contour level corresponds to 1\,mJy/beam for NVSS image and 
0.4\,mJy/beam for the FIRST image. Crosses indicate the position of the
galactic nucleus. Lower panel: SDSS spectrum of SDSS\,J144532.17+350253.5.
Based on the appearance of this spectrum, SDSS\,J144532.17+350253.5 is a normal 
galaxy, although it must have been active and RL before.}
\label{144532+350253}
\end{figure*}

Obviously, the later the evolutionary stage, the more advanced the decay of 
relics as shown in the examples in Figs.\,\ref{152719+105015} 
and\,\ref{155704+340309}. Here, the lobes are even more dispersed compared 
to those in Figs.\,\ref{082904+571541} and\,\ref{123915+531416}. This shift 
of the radio appearance has, in turn, a counterpart in the optical domain 
where the signatures of activity have faded out nearly completely: the 
continuum is typical for normal galaxies, broad lines have disappeared, and 
only single narrow lines are present. A further assumption that could be made 
is that the inclusion of the whole time spent in the intermediate state to the 
above time budget would still allow for it to remain within the $10^8$\,yr 
limit. If this is correct, galaxies should exist that show no signatures of
activity in their optical spectra at all but are straddled with extremely weak
relic radio lobes. Based on the examples shown in Figs.\,\ref{130526+365520} 
and\,\ref{144532+350253}, this could be the case. The SDSS spectra of 
\object{SDSS\,J130526.77+365521.0} and \object{SDSS\,J144532.17+350253.5} 
indicate that they are normal galaxies, yet are accompanied by radio relics, 
albeit diffuse to such an extent that VLA in B-conf. was unable to render them 
in FIRST.

The extinction of activity in galaxies is certainly 
not limited to those listed in Table\ref{gal}. To date, perhaps the most 
spectacular evidence of a past activity in a galaxy that currently appears 
as normal was brought by \citet{Schaw2010} for \object{IC\,2497}. No more 
than 70,000 years ago, it hosted a luminous quasar, the light echo of which 
is now observable in the form of a nebulous object known as Hanny's 
Voorwerp, located 15-25\,kpc (in projection) from IC\,2497. The authors 
point out that the ``death'' of the QSO hosted by IC\,2497 must have been 
sudden so that, statistically, objects similar to Hanny's Voorwerp should be
very rare because of their ephemeral nature. Two objects in our sample -- 
SDSS\,J082905.01+571541.6 and SDSS\,J123915.40+531414.6 shown in 
Figs.\,\ref{082904+571541} and\,\ref{123915+531416}, respectively -- are 
also rare and accordingly likely to remain in a short, transitional phase.
It can be speculated that this is earlier than the current evolutionary stage of
IC\,2497. On the other hand, the remaining four presented here in some 
detail are clearly post-active but, unlike for IC\,2497, radio relics instead
of optical relics are the protagonists of their past active episode.

\section{Summary}

We attempted to show an evolutionary path for galactic nuclei
analogous to that of XRBs between VH state and quiescence. The presence of 
radio lobes inflated in the VH state but later gradually dispersing and fading 
out makes it possible to discern three stages of the evolution of galactic 
nuclei that follow the VH state: the HS, intermediate, and quiescence. The
switch to the HS state was covered in Paper\,I. In this work, optical and radio
signatures of the two remaining stages have been pointed out.

\begin{acknowledgements}

\item This research has made use of the NASA/IPAC Extragalactic Database 
(NED), which is operated by the Jet Propulsion Laboratory, California 
Institute of Technology, under contract with the National Aeronautics and 
Space Administration.

\item Funding for the SDSS and SDSS-II has been provided by the Alfred P. 
Sloan Foundation, the Participating Institutions, the National Science 
Foundation, the U.S. Department of Energy, the National Aeronautics and 
Space Administration, the Japanese Monbukagakusho, the Max Planck Society, 
and the Higher Education Funding Council for England. The SDSS Web Site is 
http://www.sdss.org/. The SDSS is managed by the Astrophysical Research 
Consortium for the Participating Institutions. The Participating 
Institutions are the American Museum of Natural History, Astrophysical 
Institute Potsdam, University of Basel, University of Cambridge, Case 
Western Reserve University, University of Chicago, Drexel University, 
Fermilab, the Institute for Advanced Study, the Japan Participation Group, 
Johns Hopkins University, the Joint Institute for Nuclear Astrophysics, the 
Kavli Institute for Particle Astrophysics and Cosmology, the Korean 
Scientist Group, the Chinese Academy of Sciences (LAMOST), Los Alamos 
National Laboratory, the Max-Planck-Institute for Astronomy (MPIA), the 
Max-Planck-Institute for Astrophysics (MPA), New Mexico State University, 
Ohio State University, University of Pittsburgh, University of Portsmouth, 
Princeton University, the United States Naval Observatory, and the 
University of Washington.

\end{acknowledgements}

\end{document}